\documentclass[10pt,leqno]{amsart}
\usepackage{graphicx}

\usepackage{indentfirst,csquotes}

\usepackage{amssymb,amsthm,amsmath}

\usepackage{multirow}
\usepackage{subfigure}
\usepackage{url}

\usepackage{xcolor,paralist,hyperref,fancyhdr}
\hypersetup{ colorlinks=true, linkcolor=black, filecolor=black, urlcolor=black }

\usepackage{lipsum}

\begin{document}

\title{\vspace{-75pt}Hi-SAM: A high-scalable authentication model \\for satellite-ground Zero-Trust system \\using mean field game\vspace{-15pt}} 

\author[aaa]{\scriptsize Xuesong Wu$^{a,}$*}
\author[aaa]{\scriptsize Tianshuai Zheng$^a$} 
\author[aaa]{\scriptsize Runfang Wu$^a$} 
\author[bbb]{\scriptsize \\Jie Ren$^b$} 
\author[ccc]{\scriptsize Junyan Guo$^c$} 
\author[aaa]{\scriptsize Ye Du$^{a,}$*}


\date{\today}

\maketitle
\let\thefootnote\relax
\footnotetext{* Corresponding author.}
\footnotetext{$^a$ School of Computer and Information Technology, Beijing Jiaotong University, Beijing, 100044, China.} 
\footnotetext{$^b$ China Assets Cybersecurity Technology Co.,Ltd., Beijing, 100043, China.}
\footnotetext{$^c$ China Telecom Cybersecurity Technology Co.,Ltd., Beijing, 100020, China.}
\footnotetext{{\it E-mail:} Xueso.Wu@aliyun.com, ydu.bjtu.edu.cn} 

\begin{abstract}\vspace{-35pt}
As more and more Internet of Thing (IoT) devices are connected to satellite networks, the Zero-Trust Architecture brings dynamic security to the satellite-ground system, while frequent authentication creates challenges for system availability. To make the system's accommodate more IoT devices, this paper proposes a high-scalable authentication model (Hi-SAM). Hi-SAM introduces the Proof-of-Work idea to authentication, which allows device to obtain the network resource  based on frequency. To optimize the frequency, mean field game is used for competition among devices, which can reduce the decision space of large-scale population games. And a dynamic time-range message authentication code is designed for security. From the test at large population scales, Hi-SAM is superior in the optimization of authentication workload and the anomaly detection efficiency. 
\end{abstract} 

\bigskip
\vspace{-15pt}
\section{Introduction}

As multi-domain networks converge, a wide variety of Internet of Thing (IoT) devices are added to the integration network. Satellite nodes have become important access points due to the flexibility and wide coverage of satellite networks. However, the increase in the number of IoT devices has increased the attack surface of the entire network, so Zero-Trust Architecture (ZTA) \cite{stafford2020zero} has been introduced to the satellite-ground system. It secures a system through continuous authentication. Compared to traditional authentication, it is more suitable for the increasing number of sensors in future scenarios. Especially for wireless networks that handover frequently, ZTA greatly enhances their dynamics.

However, while ZTA brings security, it also poses a challenge to the availability of the system. The specifics are as follows: 

\noindent(1) Constrained resources. Integration networks will absorb more and more IoT devices. Because of the expansion of IoT devices, frequent authentication can expose satellites, as access points, to shorter lifetimes and higher stress. 

\noindent(2) Diverse demands. The variety of IoT devices is growing, and not all IoT devices require frequent authentication, such as some environmental sensors. Excessive security requirements for the system will only make these devices less energy efficient.

\noindent(3) Static authentication framework: The vast majority of the work \cite{DBLP:journals/tc/LiuHRXZCM23, DBLP:journals/tifs/GeZ24, DBLP:journals/compsec/ShahSSABD21} on ZTA is the design to simplify the authentication process or reduce the computational cost. These are optimizations for static authentication, and the frequency does not change. As the number of devices increases, a fixed frequency can quickly bottleneck the system.

\noindent This severely limits the scalability of satellites. To reduce the frequency while still maintaining security, a device-differential authentication is necessary.

For above motives, the most obvious difference among the devices is their demand for resources. Compared with low-demand devices, the high-demand have large traffic and little concern for energy and pose a greater threat when compromised. To maintain security, high-demand devices need to be authenticated more frequently. Based on the Proof-of-Work idea \cite{DBLP:journals/mansci/LiRS24}, the authentication is used as workload to compete for resources and a High-Scalable Authentication Model (Hi-SAM) is proposed. 

Hi-SAM consists of two parts. (1) In order for all devices to recognize a share of the gains from the competition, which means that the game converges. One is a first-order mean field game (MFG) \cite{lasry2007mean,caines2021mean, DBLP:journals/wcl/ChenLHT24} designed for authentication frequency optimization in a large-scale population. The idea of mean field avoids decision space explosion in a game when the agents' number trend to be infinite. (2) The other is a message authentication code (MAC) \cite{DBLP:conf/wisec/0003BH22} based dynamic time-range authentication policy. Due to the idea of the Proof-of-Work, IoT devices have different authentication frequencies. And the low-demand devices may be sleeping for too long, so the authentication set a dynamic time-range MAC with a penalty policy for oversleeping to incentivize low-demand devices. This MAC changes the message based on the sleeping time to secure the device-differential authentication.

The main contributions of this paper are as follows:

\noindent (1) Introduce the Proof-of-Work idea into ZTA so that the authentication frequency is adjusted based on devices' threat level and Hi-SAM's scalability can be greatly enhanced;

\noindent (2) Design a population frequency optimization based on MFG and propose a triangular density estimation method for the first-order mean field;

\noindent (3) Design a dynamic time-range MAC with a penalty policy. It utilizes dynamic bit-shift to secure authentication at different frequencies. And oversleeping devices will be penalized in the workload.

The remainder is organized as below. Section II is the related works. Section III is the system formulate. Section IV introduces Hi-SAM. Numerical results are shown in Section V. Section VI is the conclusion.
\vspace{-10pt}
\section{Related Works}
In the satellite-ground integration network, the authentication is a necessary function for system security. Due to the lack of dynamism in traditional authentication methods, the next-generation network focuses on ZTA. 

In the studies of ZTA, the implementation and optimization of authentication are the popular directions for research. Chen et al. \cite{DBLP:conf/hpcc/ChenSS21} proposed a multi-level two-way authentication for the mobile Internet. This method makes the authentication process more flexible. Tang et al. \cite{tang2023privacy} proposed a privacy-preserving authentication scheme for ZTA. For the practicability, it applied a single packet authentication. Shah et al. \cite{DBLP:journals/compsec/ShahSSABD21} a lightweight continuous device-to-device authentication. It used channel state information to generate session keys to reduce the complexity. Liu et al. \cite{DBLP:journals/tc/LiuHRXZCM23} used blockchain to design a IoT-ZTA for information sharing. However, the above approaches improve the authentication process and reduce the authentication complexity, which means that a fixed frequency of authentication can make the network stress increase dramatically when IoT devices increase. Moreover, they considered scenarios that do not distinguish between IoT devices. This is not practical in a integration network and makes the device that originally had very little business traffic need to spend more energy on security authentication. This is an unacceptable cost for them. To enlarge the scalability of network, Hi-SAM focuses on the difference among devices and uses Proof-of-Work to design a device-differential authentication model. By binding resource allocation and authentication frequency, IoT devices can determine their own workload in Hi-SAM.

To gain resources, IoT devices need to compete with each other. Intense competition stresses the system, while negative competition decreases the time to anomaly detection. Both scenarios are undesirable. Therefore, a reasonable and efficient game among devices is necessary for Hi-SAM. Ge et al. \cite{DBLP:journals/tifs/GeZ24} used a dynamic Markov game model to defense against the lateral movement attack. However, the game model assumed a small set of actions. When there are as many devices as there are in practice, its decision space explodes, making it impossible to take actions effectively. Wang et al. \cite{DBLP:journals/twc/WangYTH14} introduced MFG to Ad hoc networks' security. This is a method that uses probability densities to replace a class of individual decisions \cite{DBLP:journals/tcst/AzizC17}. When the population size tends to infinity, its mean field is able to significantly reduce the decision space. In addition, Li et al. \cite{DBLP:journals/mansci/LiRS24} also used MFG to optimize the cryptocurrency mining. Therefore, Hi-SAM select MFG to achieve the device-differential authentication frequency optimization.

With above problems, a MFG based high-scalable authentication model for satellite-ground Zero-Trust system is designed in this paper. Under the model of MFG, IoT devices can efficiently optimize the frequency of authentication, while the frequency of population creates a tradeoff between security and availability. Furthermore, since the security of conventional authentication is degraded by the difference in frequency and the oversleeping brought by MFG, this paper designs a MAC based dynamic time-range authentication for this purpose.

\section{System Formulation}

In this satellite-ground Zero-Trust system, the authentication number (workload) of a IoT device can be exchanged for resources, such as reputation, bandwidth, and QoS. A description of the system model will be given in the following.

\subsection{Access Point}
The access point, a satellite, acts as a manager, and it needs to allocate resource and make policy. It is defined by a triple

\vspace{-2pt}
\begin{equation}
  \abovedisplayshortskip=-4pt
  \label{model_1}
    AP:=<N,\ F_P,\ \pi>,\nonumber
\end{equation}

\noindent where $N$ is the number of population, $F_P$ is the maximum population authentication frequency within a time unit $T$, and $\pi$ is the ZTA policy. Among them,  $F_P$ is a fixed parameter which is determined by $AP$'s performance. As can be seen, if the population size $N$ is to be expanded, the policy $\pi$ must make concessions on security because $F_P$ can not be changed. 

\subsection{Terminal}
The terminal, or the user equipment, acts as a member, and it has to compete for network resources and respond to the policy $\pi$. It can be represented as a binary
\vspace{-2pt}
\begin{equation}
  \abovedisplayshortskip=-4pt
  \label{model_2}
    UE:=<r,\ \alpha>,\nonumber
\end{equation}
where $d$ is the expected demand, and $\alpha$ is the individual's authentication frequency. Under limit $F_P$ by $AP$,  $\alpha \le \min\{F_I,F_P/N\}$, where $F_I$ is maximum individual frequency and indicates the response time of abnormal events. $F_I$ is set according to the security requirements. Then, the individual's work $x$ can be expressed by a differential equation
\vspace{-2pt}
\begin{equation}
  \abovedisplayshortskip=-4pt
  \label{model_3}
    \text{d}x=\alpha \ \text{d}t.
\end{equation}
With the Proof-of-Work idea, the allocated resources $\hat d$ is 
\vspace{-2pt}
\begin{equation}
  \abovedisplayshortskip=-4pt
  \label{model_4}
    \hat{r}=R\cdot\frac{\alpha}{X/T},
\end{equation}
where $R=\sum_{i=1}^{N} r_i$ is total resource of system and $X=\sum_{i=1}^{N} x_i^T$ ($x_i$ at time $T$) is the workload of the population. It can be seen that $\hat r$ is allocated according to work. From eq. (\ref{model_4}), the conflict in a population comes to the fore. The more individual works, the more resources can be occupied. However, the more population works, the less valuable a work is. Hence, a loss function for the game is constructed as
\vspace{-2pt}
\begin{equation}
  \abovedisplayshortskip=-4pt
  \label{model_5}
    l(r,\alpha,X)=\frac{1}{F_P-X/T}\cdot\alpha+\frac{Xr}{RT}\cdot\frac{1}{\alpha}.
\end{equation}
eq. (\ref{model_5}) is divided into two parts by the plus sign. The left part indicates the competitive pressure. The population frequency $X/T$ is closer to the maximum frequency $F_P$, the loss ratio of $\alpha$ is larger. The right part indicates the inverse of the individual reward. The $X/(XT)$ is closer to $1$, the more reward can be obtained. With the loss function and the policy $\pi$, $UE$ can determine a suitable frequency $\alpha$ . 

\subsection{Framework of Authentication}
\begin{figure*}[h]
  \vspace{-8pt}
  \centerline{\includegraphics[width=4.5in]{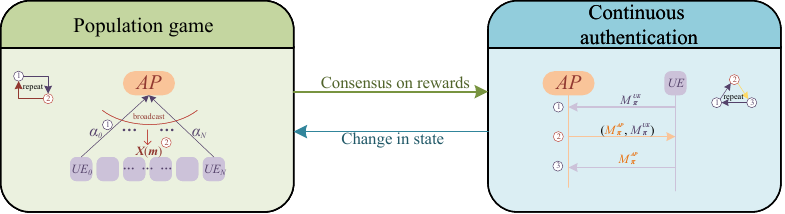}}
  \vspace{-4pt}
  \caption{The framework of Hi-SAM.}
  \label{frame}
  \vspace{-10pt}
\end{figure*}

From the description of $AP$ and $UE$, the framework of Hi-SAM has two main parts shown in Fig. \ref{frame}. 

Firstly, the population negotiates workload under a MFG. $AP$ broadcasts the total resources $D$, the acceptable individual frequency $F_m$, and the estimated population workload $X(m)$ as the default information. During this period, $UE$s calculate the frequency $\alpha$ based on the game with eq. (\ref{model_5}). Then, $AP$ keeps counting frequency $\alpha$ and uses the mean field to estimate the population's workload $X(m)$ based on a distribution $m$. Once $X(m)$ is calculated, it will be broadcast to $UE$s. This process is repeated until the competition is agreed upon, and the system task will be converted to continuous authentication

Afterwards, $UE$s have to repeat continuous authentication according to the policy $\pi$ and the consensus of workload $\alpha$. Considering that the dynamic game makes the degree of activity varies enormously, a dynamic time-range MAC for $\pi$ is needed to ensure the security in different time scale. When the change of system's state occurs, such as changes in individual demands or members, the continuous authentication will be converted to the task of population game.

\section{The High-Scalable Authentication Model}
\subsection{A MFG based Authentication Frequency Optimization}
MFG can avoid decision space explosion when population size is large. And as shown in Fig. \ref{frame}, the game can be divided into two steps, individual decisions (optimal control) and population trends (mean field theory). To build a game, the following work is necessary. These works cover optimal control and field theory. It is beyond the purpose of this paper to explain in detail these pre-existing findings which can be read in \cite{lasry2007mean,caines2021mean, DBLP:journals/wcl/ChenLHT24, note}.
\subsubsection{Individual Decisions}
This part is deployed on $UE$, which is an optimal control on $\alpha$. Since most of steps involve only individual states, the equations are in scalar form. 

Firstly, to downscale the decision space, the population's workload $X$ need be calculated with a distribution of all individuals' workload $m(t)=\prod_{i=1}^N m_i(t):[0,T]\rightarrow \boldsymbol{P} (\mathbb{R}^N)$ . Then, eq. (\ref{model_5}) can be rewritten as
\vspace{-2pt}
\begin{equation}
  \abovedisplayshortskip=-4pt
  \label{MFG_1}
    l(r,\alpha,m(T))=\frac{1}{F_P-X(m)/T}\cdot\alpha+\frac{X(m)r}{RT}\cdot\frac{1}{\alpha},
\end{equation}
where $X(m)=\int_0^{F_mT}\sum_{i=1}^{N}{x_i}\cdot m(T)\ \text{d}\boldsymbol{x}=\sum_{i=1}^{N}(\int_0^{F_mT}x_i\cdot m_i(T)\text{d}x_i)$ (vector form) is the expectation of population workload and $t=T$ is the focus point. So, a system payoff function \cite{note} is denoted as
\vspace{-2pt}
\begin{equation}
  \abovedisplayshortskip=-4pt
  \label{MFG_2}
    J(t,x,\alpha)=\mathbb{E}\Big[\int_{t}^{T}  l (r,\alpha,m(T))\ \text{d}s+G(x,T)\Big],
\end{equation}
where $G(x,T)$ is a terminal cost and defined as
\vspace{-2pt}
\begin{equation}
  \abovedisplayshortskip=-4pt
  \label{MFG_3}
    G(x,T)=\frac{x_T}{F_mT}-1,
\end{equation}
where $x_T$ is the workload at $T$ and $F_m=\min\{F_I,F_P/N\}$. And the value function of system is
\vspace{-2pt}
\begin{equation}
  \abovedisplayshortskip=-4pt
  \label{MFG_4}
    v(t,x)=\inf_\alpha J(t,x,\alpha).
\end{equation}\
The Hamilton-Jacobi equation \cite{note} with $v$ is 
\vspace{-2pt}
\begin{equation}
  \abovedisplayshortskip=-4pt
  \label{MFG_5}
  \begin{aligned}
    \begin{cases}
      -\partial_tv(t,x)+H(t,x,Dv(t,x),m(T))=0,\\
      v(T,x)=G(x,T),
      \end{cases}
  \end{aligned}
\end{equation}
where $H$ is the Hamiltonian. The optimal feedback of each individual can be calculated as
\begin{equation}
  \abovedisplayshortskip=-4pt
  \label{MFG_7}
    \alpha^*(t)=\sqrt{\frac{1}{\mu_2(\partial_x v + \mu_1)}},
\end{equation}
where $\mu_1={1}/({F_P-X(m)/T})$ and $\mu_2={DT}/({X(m)r})$. Under the optimal feedback, the infimum of $H$ is
\begin{equation}
  \abovedisplayshortskip=-4pt
  \label{MFG_8}
    H^*=2\sqrt{\frac{\partial_x v + \mu_1}{\mu_2}}.
\end{equation}
With the terminal cost $G(x,T)$, $v$ can be obtained by partial differential equation eq.(\ref{MFG_5})
\begin{equation}
  \abovedisplayshortskip=-4pt
  \label{MFG_9}
    v(t,x)=\frac{x}{F_mT}-1+2\sqrt{\frac{\mu_1+1/F_mT}{\mu_2}}(T-t).
\end{equation}
\newtheorem{remark}{\bf Remark}
\begin{remark}
  The value function with the optimal feedback can also be computed directly by its definition
  \begin{equation}
    \abovedisplayshortskip=-4pt
    \label{MFG_a}
    \begin{aligned}
      v(t,x)&=\inf_\alpha \Big\{\mathbb{E}\Big[\int_{t}^{T}  l (r,\alpha,m(T))\ \text{d}s+G(x,T)\Big]\Big\}\\
      &=\inf_\alpha \Big\{\big((\frac{1}{F_mT}+\mu_1)  \alpha +\frac{1}{\mu_2\alpha}\big)(T-t)\hspace{-2pt}+\hspace{-2pt}\frac{\alpha t}{F_mT}\hspace{-2pt}-\hspace{-2pt}1\Big\}\\
      &\ge 2\sqrt{\frac{\mu_1+1/F_mT}{\mu_2}}(T-t)+\frac{x}{F_mT}-1.\nonumber 
    \end{aligned}
  \end{equation}
  And $\alpha^*$ in eq. (\ref{MFG_10}) holds the equality sign. \hfill \qed
\end{remark}
Finally, the optimal frequency is
\begin{equation}
  \abovedisplayshortskip=-4pt
  \label{MFG_10}
    \alpha^*=\sqrt{\frac{1}{\mu_2(\mu_1+1/(F_mT))}},
\end{equation}
and $UE$ delivers the optimal frequency $\alpha^*$ to $AP$ once the computation is complete.

\subsubsection{Population Trends}

This part is deployed on $AP$, which calculates a mean field $m$. Since most of steps involve all individuals' states, equations are in vector form. 

\begin{figure*}[h]
  \vspace{-8pt}
  \centerline{\includegraphics[width=4in]{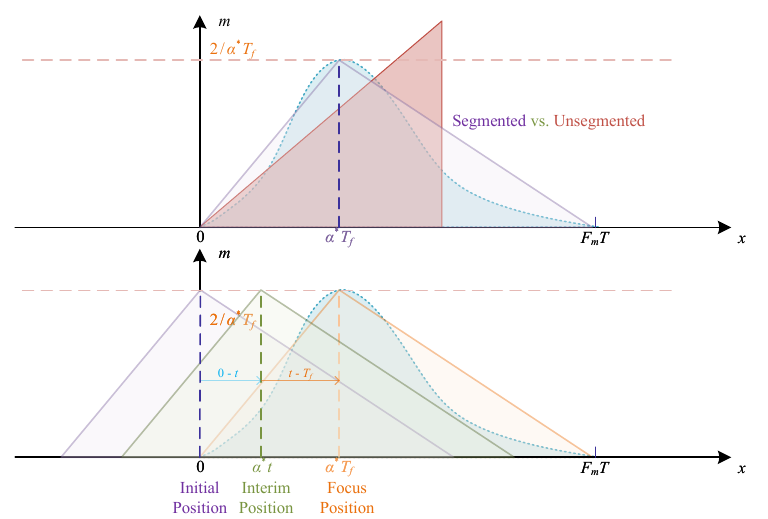}}
  \vspace{-8pt}
  \caption{Example of the triangular density distribution.}
  \vspace{-8pt}
  \label{mfg}
\end{figure*}

When $AP$ receives all $\boldsymbol{\alpha}^*$, it begins to evaluate the workload density. Based on the Fokker-Planck equation \cite{note}, there is
\begin{equation}
  \abovedisplayshortskip=-4pt
  \label{MFG_11}
    \partial_tm(t,\boldsymbol{x})+\text{div}[\boldsymbol{\alpha}^*\cdot m(t,\boldsymbol{x})]=0.
\end{equation}
Clearly, eq. (\ref{MFG_11}) degenerates to a continuity equation. Considering that its characteristic lines are linear, if they are used to evaluate the density, a `distorted' shown as a red right angled triangle in the upper part of Fig. \ref{mfg} will occurs. To reduce the distortion, $m(t,\boldsymbol{x})$ is set as a segmented function, and a violet triangular region better fits the real distribution. The constructing steps $m(t,\boldsymbol{x})$ are described next.

Let $x_m=F_m\cdot T$ , $\boldsymbol{x}^*=\boldsymbol{\alpha}^*\cdot T_f$ , and  $x_b$ be the length of the triangle base, a more general form of $m_i$ is given,  
\begin{equation}
  \abovedisplayshortskip=-4pt
  \label{MFG_12}
  \begin{aligned}
      m_i(T_f, t,x_i)=
      \hspace{-5pt}\begin{cases}
        \max\{0, \frac{2\big(x_i-\alpha^*_i(t-T_f)-(x_i^*-\gamma x_b)\big)}{x_b\gamma} x_b\},\hfil &x_i<\alpha_i^*t,\\
        \max\{0, -\frac{2\big(x_i-\alpha_i^*(t-T_f)-(x_i^*+(1-\gamma)x_b)\big)}{x_b(1-\gamma)x_b}\},\hfil&x_i\ge\alpha_i^*t,
        \end{cases}
  \end{aligned}
\end{equation}
where $\gamma\in(0,1)$ is the density weight of $x_i<x_i^*$, $T_f\in(0,T]$ is a target moment which is focused on. (And there is some abuse of notation that the multiplications and inverses are all for elements in the vector.)

To satisfy $\int m(t,\boldsymbol{x})\ \text{d}  \boldsymbol{x}=1$, this density distribution function, shown in Fig. \ref{mfg}, has $x^{\max}-x^{\min}=x_b$. Then, the height of the triangle is fixed as $2/x_b$. And the slope of the line is $2/(x_b{\gamma}  x_b)$  (or $-2/[x_b(1-{\gamma})x_b)]$). The coordinate of the vertex is $x_i^v=\alpha_i^*t$ , and it also the division point of the segmented function. Using the vertex, the initial condition can be computed, $m_i(T_f,0,x_i)=m_i(T_f,0,0)=2/x_b$.

Up to now, a triangular density distribution is designed. A more intuitive understanding of $m_i(T_f, t, x_i)$ is shown in the bottom half of Fig. \ref{mfg}. It can be viewed as a triangular distribution based on the moment $T_f$ and moving on the $x$-axis as $t$ changes. Its $x^*_i$ is similar to the mean and $x_b$ is similar to the variance.

For mathematical simplicity, this paper unitizes the time unit $T=1$, $x_b=x_m$ , ${\gamma} x_b=x^*$, and $T_f=T=1$. Therefore, the density distribution function is
\begin{equation}
  \abovedisplayshortskip=-4pt
  \label{MFG_13}
  \begin{aligned}
    m(t,\boldsymbol{x})=\begin{cases}
      \prod_{i=1}^N\Big(\max\{0, \frac{2\big(x_i-\alpha_i^*(t-1)\big)}{F_m\alpha_i^*}\}\Big),\qquad x_i<\alpha_i^*t,\\
      \prod_{i=1}^N\Big(\max\{0, -\frac{2\big(x_i-F_m-\alpha_i^*(t-1)\big)}{F_m(F_m-\alpha_i^*)}\}\Big),\qquad x_i\ge\alpha_i^*t.
      \end{cases}
  \end{aligned}
\end{equation}

Finally, $AP$ evaluates the workload $X(m)$ at $t=1$ by
\vspace{-4pt}
\begin{equation}
  \abovedisplayshortskip=-4pt
  \label{MFG_14}
  \begin{aligned}
    X(m)&=\int_0^{\boldsymbol{\alpha}^*}\sum_{i=1}^{N}{x_i}\cdot m(T,\boldsymbol{x})\text{d}\boldsymbol{x}+\int_{\boldsymbol{\alpha}^*}^{F_m}\sum_{i=1}^{N}x_i\cdot m(T,\boldsymbol{x})\text{d}\boldsymbol{x}\\
    &=\sum_{i=1}^{N}\Big(\int_0^{{\alpha}_i^*}{x_i}\cdot m_i(T,{x_i})\text{d}{x_i}+\int_{{\alpha_i}^*}^{F_m}x_i\cdot m_i(T,{x_i})\text{d}{x_i}\Big)\\
    &=\sum_{i=1}^{N}\frac{\alpha_i^*+F_m}{3}.
  \end{aligned}
\end{equation}
Soon, the result $X(m)$ will be broadcast to $UE$s.

\subsubsection{Equilibrium characterization}\label{equilibrium}
The equilibrium relates to the convergence of a game model. Although, the equilibrium characteristics of the MFG does not have a fully recognized result so far, especially for first-order models \cite{DBLP:journals/nhm/GomesMT20}, and proving existence and uniqueness of an equilibrium in a general case is beyond the scope of this paper, we still give a proof for this special application scenario.

First, the population workload can be approximated as
\begin{equation}
  \abovedisplayshortskip=-4pt
  \label{MFG_15}
  \begin{aligned}
    X&=\int  \sum_{i=1}^{N}{x_i}\cdot m(\boldsymbol{x},t) \text{d} \boldsymbol{x}\\
    &\approx T\sum_{i=i}^{N}\alpha_i^*=
T\sum_{i=i}^{N}\sqrt{\frac{1}{\mu_2^i(\mu_1^i+1/(F_mT))}}.
  \end{aligned}
\end{equation}
Further, an aggregated iterative equation can be obtained
\begin{equation}
  \abovedisplayshortskip=-4pt
  \label{MFG_16}
  \begin{aligned}
    X &\approx T\sum_{i=1}^N\sqrt{\frac{(F_PT-X)Xr_i}{T^2R}}\\
    \Rightarrow X&=(F_PT- \frac{R}{(\sum_{i=i}^{N}\sqrt{r_i})^2}\cdot X).
    \end{aligned}
\end{equation}
Due to  $1/N\le R/(\sum_i\sqrt{r_i})^2 < 1$ , let the right side of eq. (\ref{MFG_16}) be $\varphi(X)$, and the following conclusion can be obtained:
\begin{itemize}
  \item \vspace{-2pt}$\varphi(X)$ is continuous on $(N,F_PT]$ , where $N$ indicates that each individual is authenticated only once;
  \item  Because of $D/(\sum_i\sqrt{d_i})^2 < 1$ , there is $\varphi(X)\in[F_PT/2,F_PT]\subseteq [N,F_PT]$ ($N\ge2$) ;
  \item $|\varphi'(X)|<1$.\vspace{-2pt}
\end{itemize}
With above conclusion, the first-order MFG in this paper satisfies the classical fixed-point theorem for numerical iterative computation. Thus, there exists an unique fixed point $X^*\in(0,F_PT]$ for the iteration in eq. (\ref{MFG_16}). Then, the population will eventually converge to the unique point. considering the individuals in reverse, $\boldsymbol{\alpha}^*$ in eq. (\ref{MFG_10}) is a variable determined by $X^*$. So, $\boldsymbol{\alpha}^*$ is unique if $X^*$ converges. Therefore, an unique fixed point for individuals and population is consistent, and the equilibrium is existent and unique.

\subsection{A MAC based Dynamic Time-Range Authentication Policy}
With the MFG, the security of authentication is degraded by the difference in frequency. Since low-demand individuals have longer sleep-interval, static authentication makes them easier to compromise. So, a MAC based dynamic time-range (DTR) policy $\pi$ is designed. 

\begin{figure}[!h]
  \vspace{-8pt}
  \centerline{\includegraphics[width=4in]{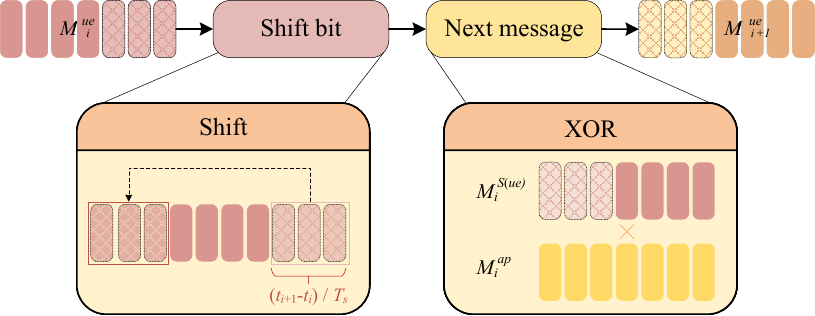}}
  \vspace{-4pt}
  \caption{The framework of dynamic time-range MAC.}
  \label{dtr}
  \vspace{-8pt}
\end{figure}

The key idea is converting sleep-interval into bit-shift. In Fig. \ref{dtr}, the bit number is calculated by a sleep-unit $T_s$ and the sleep interval. In addition, as long as the MAC is successfully sent and received, both $UE$ and $AP$ will do the same bit-shift on the message. The detail of authentication is as follows.

In this system, it is assumed that $M_0^{ap}$, $M_0^{ue}$ and $K$ , which are authentication messages and key, have been agreed upon in the registration phase. First, $UE$ calculates the shift-bit
\vspace{-4pt}
\begin{equation}
  \abovedisplayshortskip=-4pt
  \label{dtr_1}
  b^s_{i+1}= \lfloor\frac{t^{ue}_{i+1}-t^{ue}_{i}}{T_s}+0.5\rfloor,
\end{equation}
and shifts the $M^{ue}_{i}$ to $M^{s(ue)}_{i}$ according to the operation in Fig. \ref{dtr}. Then, using  $M^{s(ue)}_{i}$ and $M^{ap}_{i}$, the authentication message $M^{ue}_{i+1}$ can be generated
\begin{equation}
  \abovedisplayshortskip=-4pt
  \label{dtr_2}
  M^{ue}_{i+1}=\text{XOR}(M_{i}^{s(ue)},M^{ap}_{i}).
\end{equation}
Afterwards, a MAC will be calculated as $\text{MAC}(M^{ue}_{i+1},K)$, then be sent to $AP$. When $AP$ receives the $M^{ue}_{i+1}$, it first evaluates the shift-bit by
\begin{equation}
  \abovedisplayshortskip=-4pt
  \label{dtr_3}
  \hat b^s_{i+1}=\lfloor\frac{t^{ap}_{i+1}-t^{ap}_{i}}{T_s}+0.5\rfloor,
\end{equation}
where $t_{i+1}^{ap}$ and $t_{i}^{ap}$ are the arrival times of MACs. Because $b^s_{i+1}\approx \hat b^s_{i+1}$, $AP$ can compute $\hat M^{ue}_{i+1}$. By checking whether the MAC value of $\hat M^{ue}_{i+1}$ and $M^{ue}_{i+1}$ are the same, $UE$ can be verified preliminary. 

Similarly, $AP$ will send $\text{MAC}(\hat M_{i+1}^{ue}||M_{i+1}^{ap},K)$ to $UE$  to prove itself. After verifying the identity of $AP$, $UE$ will reply $\text{MAC}(\hat M_{i+1}^{ap},K)$. If $\text{MAC}(\hat M_{i+1}^{ap},K)$ $=$ $\text{MAC}(M_{i+1}^{ap},K)$, $AP$ can fully trust $UE$. Then, both parties update $M_{i+1}^{ap}$, $M_{i+1}^{ue}$ and the arrival time in preparation for the next authentication.

Back to the Proof-of-Work, the sleep interval will not affect the active individuals whose $\alpha>1/T_s$. But, negative individuals will be asleep for too long. So, a punishment mechanism is introduced: a maximum sleep period is set. If individuals' sleep period $P_s$ exceeds the limit of oversleeping $n\cdot T_s$, their workload is subtracted $\lceil P_s/T_s\rceil-n$. If the device is asleep or offline for an extended period of time so that its workload $x_i$ is less than zero, it will will be kicked out of the membership and denied access.

In this way, negative individuals maintain a certain motivation level in order to get enough resources, which allows the system to detect anomalous events in a more timely manner. Furthermore, because loss or interference can not change the interval $t_{i+1}-t_i$, it also improves system reliability.

\section{Numerical Results}

The performance of Hi-SAM is evaluated for different demand distributions and population sizes. To our best knowledge, this is the first work to discuss the dynamic ZTA's authentication frequency, so the design of simulation has no work that can be referenced. Nevertheless, Hi-SAM is compared against fixed (high: maximum frequency $\alpha=F_m/T$, and low: with DTR $\alpha=F_m/(2T)$) and demand-driven ($\alpha=F_mr/r_{max}$) frequency. And a second-authentication-frequency is considered by this system.

\begin{table}[h]
  \vspace{-12pt}
  \renewcommand{\arraystretch}{1.2}
  \caption{Parameters for simulation\label{tab:table1}}
  \vspace{-8pt}
  \centering
  \begin{tabular}[*]{l|l|l|l} 
    \hline\hline
    \hspace{-4pt}Parameter & \hspace{-4pt}Value & \hspace{-4pt}Parameter & \hspace{-4pt}Value \\ \hline
        \hspace{-4pt}Population\hspace{-1pt} size $N$ & \hspace{-4pt}{\scriptsize\{20,\hspace{-1pt} 60,\hspace{-1pt} 100\hspace{-1pt} (default),\hspace{-1pt} 140,\hspace{-1pt} 180\}} & \hspace{-4pt}$F_P$ {\scriptsize($AP$)} & \hspace{-4pt}2000 \\ \hline
        \hspace{-4pt}Demand\hspace{-1pt} distribution & \hspace{-4pt}Gaussian, $r \in (0, 20)$ &\hspace{-4pt}$F_I$  {\scriptsize($UE$)} & \hspace{-4pt}$ 20$\\ \hline
        \hspace{-4pt}Mean\hspace{-1pt} of\hspace{-1pt} demand & \hspace{-4pt}{\scriptsize\{4, 8, 10 (default), 12, 16\}} &\hspace{-4pt}Time unit {\scriptsize (T)} & \hspace{-4pt}10 s \\ \hline
        \hspace{-4pt}Variance\hspace{-1pt} of\hspace{-1pt} demand & \hspace{-4pt}{\scriptsize\{1, 2, 3 (default), 4, 5\}} &\hspace{-4pt}Sleep period & \hspace{-4pt} {$2\frac{T}{F_m}$}  \\ \hline\hline
    \end{tabular}
  \vspace{-14pt}
\end{table}
Initially, all devices are considered curious and are authenticated by $80\%$ of $F_m$. Others parameters are listed in Table \ref{tab:table1}. When one parameter is changed, the others use the default values. And to increase the confidence, the results are the average of 10 sets of randomized data.

\subsection{Numerical Convergence}\label{convergence}
Table \ref{tab:table2} lists the errors for run under different population sizes. The errors are observed to converge toward zero, and the convergence rate of iteration is greater than $0.5$. The observed convergence proves that the equilibrium of the numerical method in \ref{equilibrium} is correct and reasonable.
\begin{table}[h]
  \vspace{-10pt}
  \caption{The error of iterations\label{tab:table2}}
  \vspace{-8pt}
  \centering
  \begin{tabular}{cccc}
  \hline\hline
  \multirow{2}{*}{Rounds} & \multicolumn{3}{c}{{ $\sum_i |x_i^{t}-x_i^{t-1}| / N$}   }                                    \\ \cline{2-4} 
                          & \multicolumn{1}{c}{$N=10$} & \multicolumn{1}{c}{$N=100$} & \multicolumn{1}{c}{$N=1000$} \\ \hline
  1                       & 5.19            & 9.18             & 11.18           \\
  2                       & 0.92            & 3.62             & 4.13            \\
  3                       & 0.02            & 0.14             & 0.18              \\
  4                       & 4.84$\times10^{-4}$            & 6.18$\times10^{-3}$            & 8.22$\times10^{-3}$              \\
  5                       & 1.15$\times10^{-5}$            & 2.63$\times10^{-4}$            & 3.76$\times10^{-4}$             \\
  6                       & 2.73$\times10^{-7}$            & 1.12$\times10^{-5}$            & 1.72$\times10^{-5}$             \\
  7                       & 6.50$\times10^{-9}$            & 4.77$\times10^{-7}$            & 7.89$\times10^{-7}$             \\
  8                       & 1.55$\times10^{-10}$           & 2.03$\times10^{-8}$            & 3.62$\times10^{-8}$             \\
  9                       & 3.68$\times10^{-12}$           & 8.65$\times10^{-10}$           & 1.66$\times10^{-9}$             \\
  10                      & 8.74$\times10^{-14}$           & 3.68$\times10^{-11}$           & 7.60$\times10^{-11}$              \\ 
  \hline\hline
  \end{tabular}
  \vspace{-14pt}
  \end{table}

\subsection{Population Loss}
\begin{figure*}[h]
  \vspace{-8pt}
\centering
\subfigure[]{\includegraphics[width=1.75in]{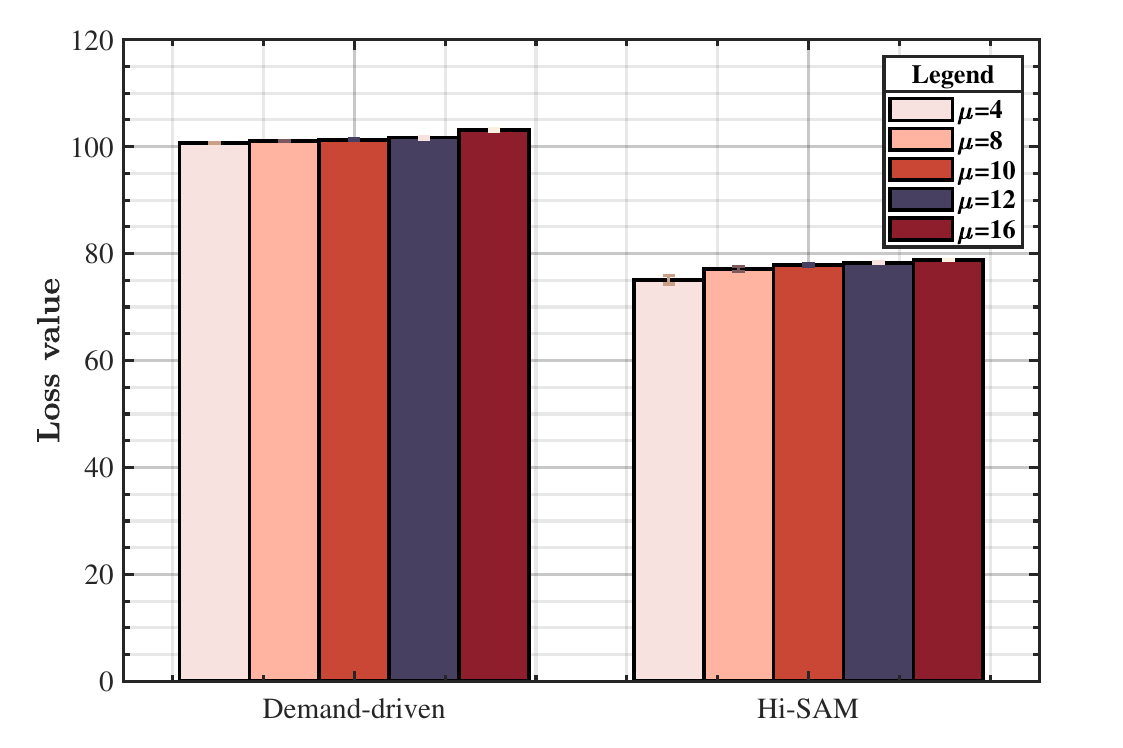}%
\label{loss-mu}}
\hfil
\subfigure[]{\includegraphics[width=1.75in]{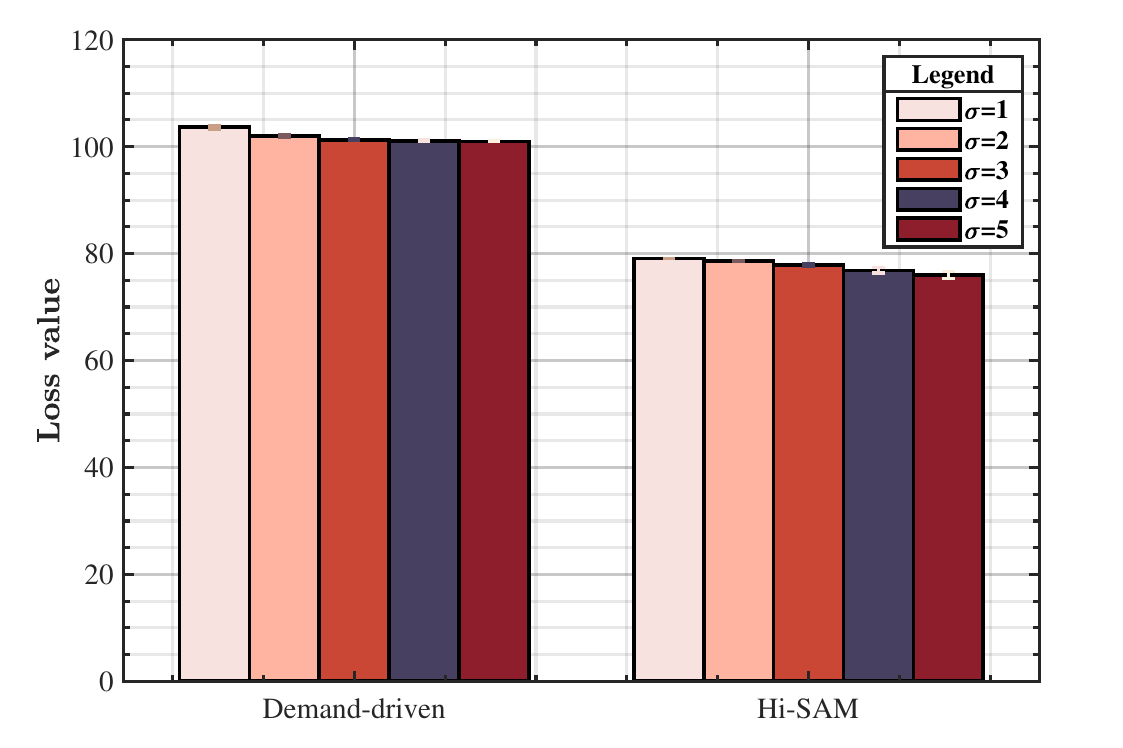}%
\label{loss-sgm}}
\hfil
\subfigure[]{\includegraphics[width=1.75in]{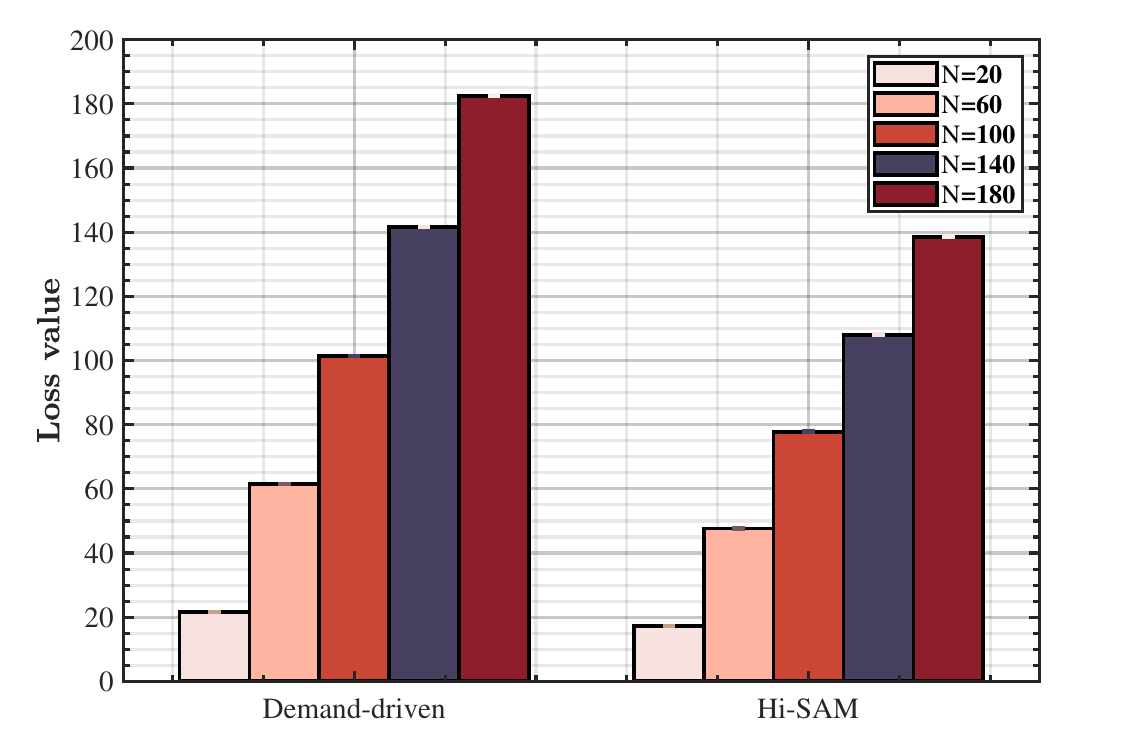}%
\label{loss-N}}
\vspace{-2pt}
\caption{Population loss under different distribution. \\(a) Change with mean. (b) Change with variance. (c) Change with size.}
\label{exp_fig_1}
\vspace{-8pt}
\end{figure*}

With the loss defined in eq.(\ref{model_5}), the demand-driven and Hi-SAM are compared under different scenarios in Fig. \ref{exp_fig_1}. From Fig. \ref{exp_fig_1}\subref{loss-mu}, both loss values rise with increasing mean of demands, and Hi-SAM has a lower cost. Fig. \ref{exp_fig_1}\subref{loss-sgm} shows the relation between loss and variance. Larger variance corresponds to smaller workloads which reflects the more consistent the demand, the more competition is incentivized. From Fig. \ref{exp_fig_1}\subref{loss-N}, as the population size increases, the loss increases as well. Compared to the demand-driven model, the loss from Hi-SAM still remains smaller.

\subsection{Anomaly Detection Time}
\begin{figure*}[h]
  \vspace{-8pt}
\centering
\subfigure[]{\includegraphics[width=1.75in]{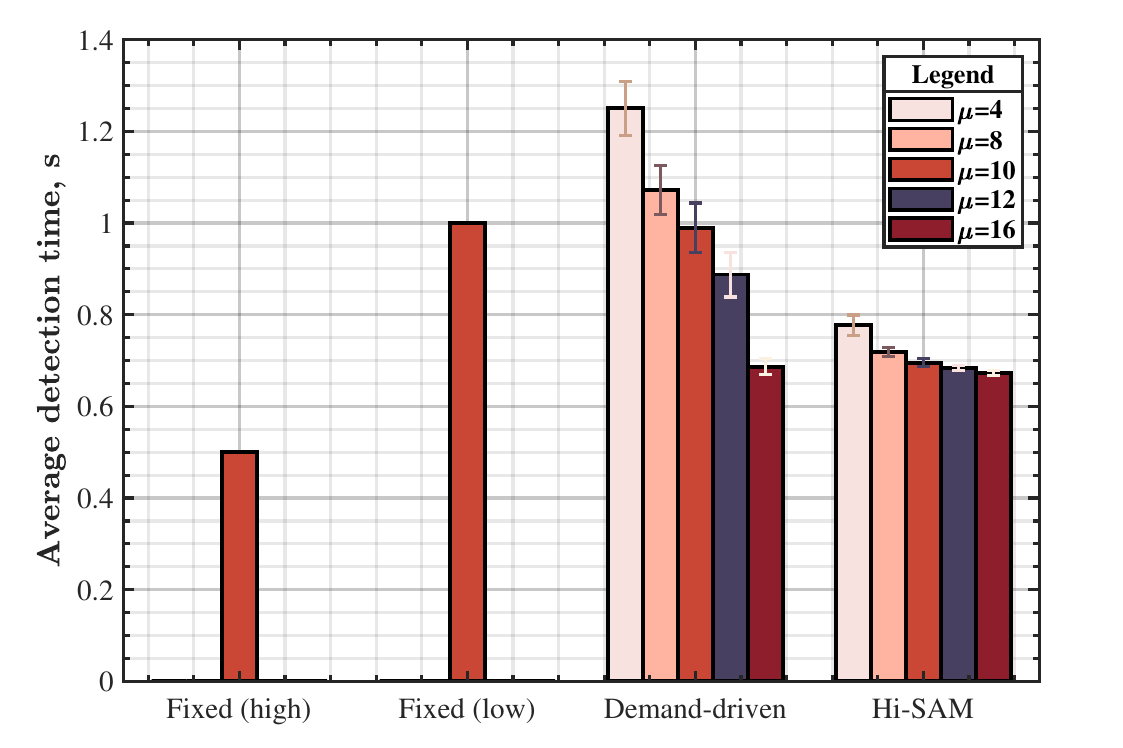}%
\label{detect-mu}}
\hfil
\subfigure[]{\includegraphics[width=1.75in]{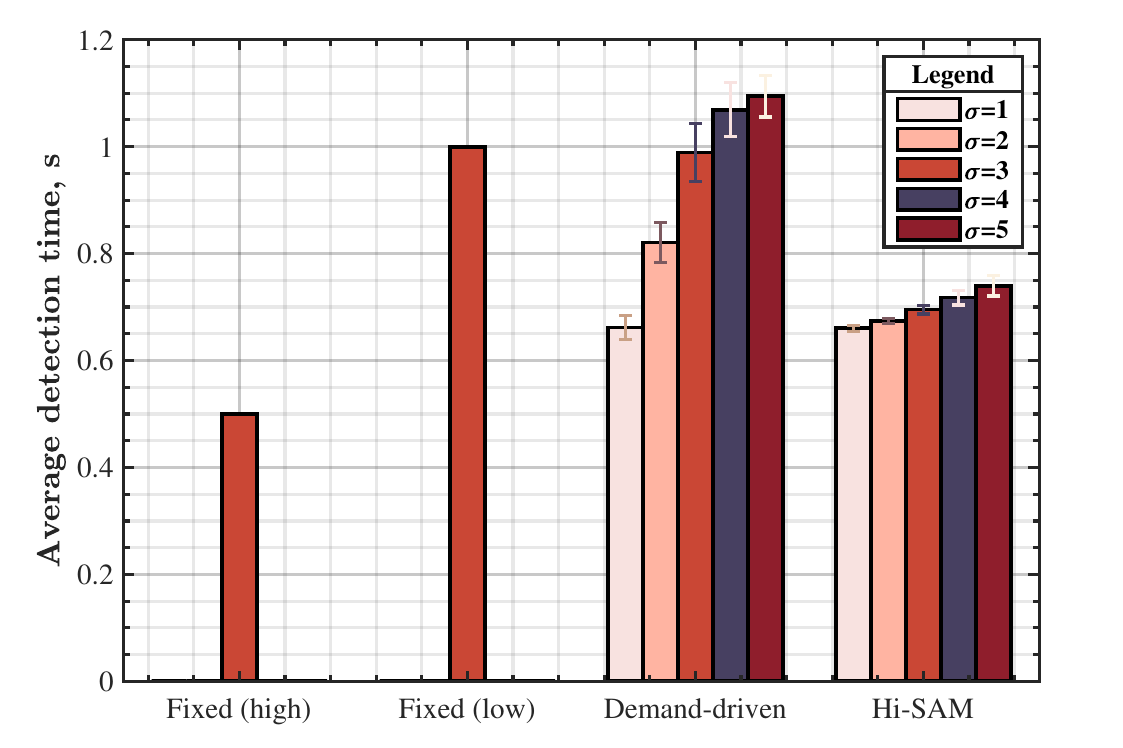}%
\label{detect-sgm}}
\hfil
\subfigure[]{\includegraphics[width=1.75in]{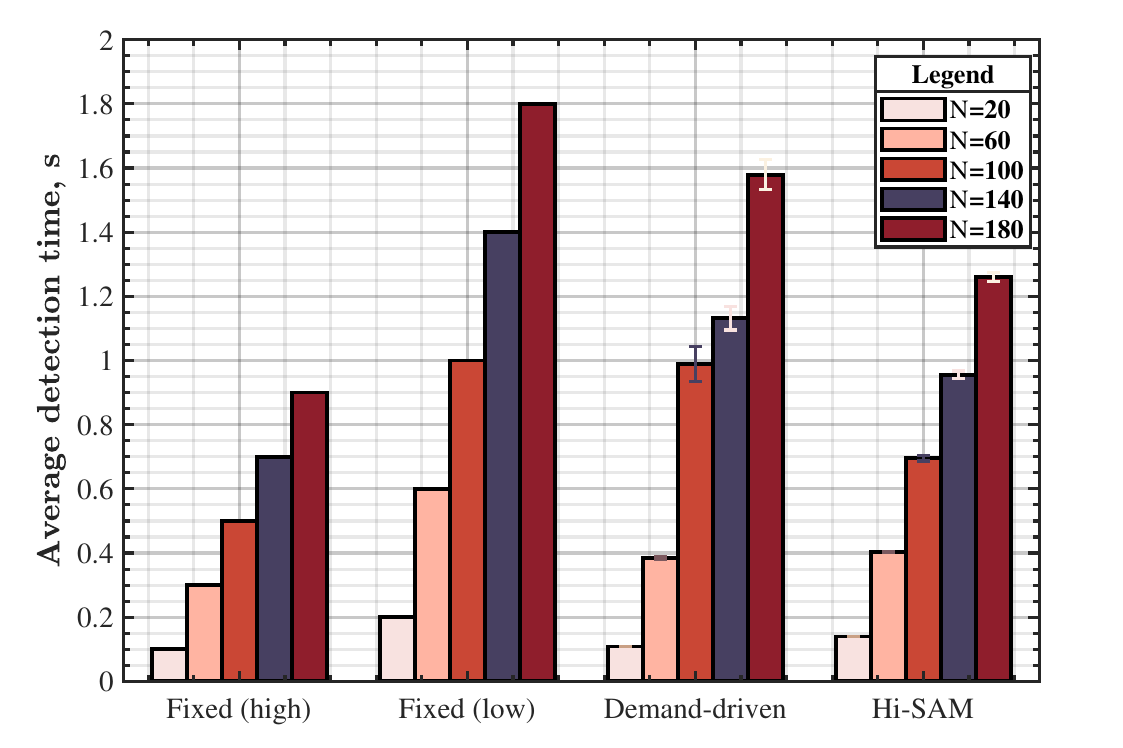}%
\label{detect-N}}
\vspace{-2pt}
\caption{Detection time under different distribution. \\(a) Change with mean. (b) Change with variance. (c) Change with size.}
\label{exp_fig_2}
\vspace{-8pt}
\end{figure*}

In this experiment, abnormal events are assumed to be random and will be detected at the next authentication. The relation between detection time and the mean and variance of demands, as well as population size, is presented in Figure \ref{exp_fig_2}. Fixed (high and low) and demand-driven frequency are used as a comparison terms.

Fig. \ref{exp_fig_2}\subref{detect-mu} shows the detection time under different mean of demands. Firstly, the fixed detection will not be affected by changes of mean. Hi-SAM has less detection time than the demand-driven model. As the mean becomes larger, the detection time is smaller, which means a higher workload and the population becomes more competitive. and the detection time of Hi-SAM keeps approaching that of the fixed high frequency.

In Fig. \ref{exp_fig_2}\subref{detect-sgm}, the detection time increases with increasing variance. As the variance increases, the demands' distribution becomes more dispersed, and competitive pressure becomes less, which makes the overall frequency lower. As a result, Hi-SAM outperforms the demand-driven model.

When the population size changes, Hi-SAM is more secure compared to the demand-driven model. In Fig. \ref{exp_fig_2}\subref{detect-N}, detection times for both are not significantly different at small sizes. However, Hi-SAM show a clear benefit at large sizes.

Overall, Hi-SAM is more superior than the demand-driven model. But that doesn't mean Hi-SAM has a higher efficiency. Next, the workload of populations will be discussed to demonstrate that Hi-SAM's high security does not result in an unacceptable spike in authentication frequency.

\subsection{Population Workload}

\begin{figure}[h]
  \vspace{-8pt}
  \centering
  \includegraphics[width=3in]{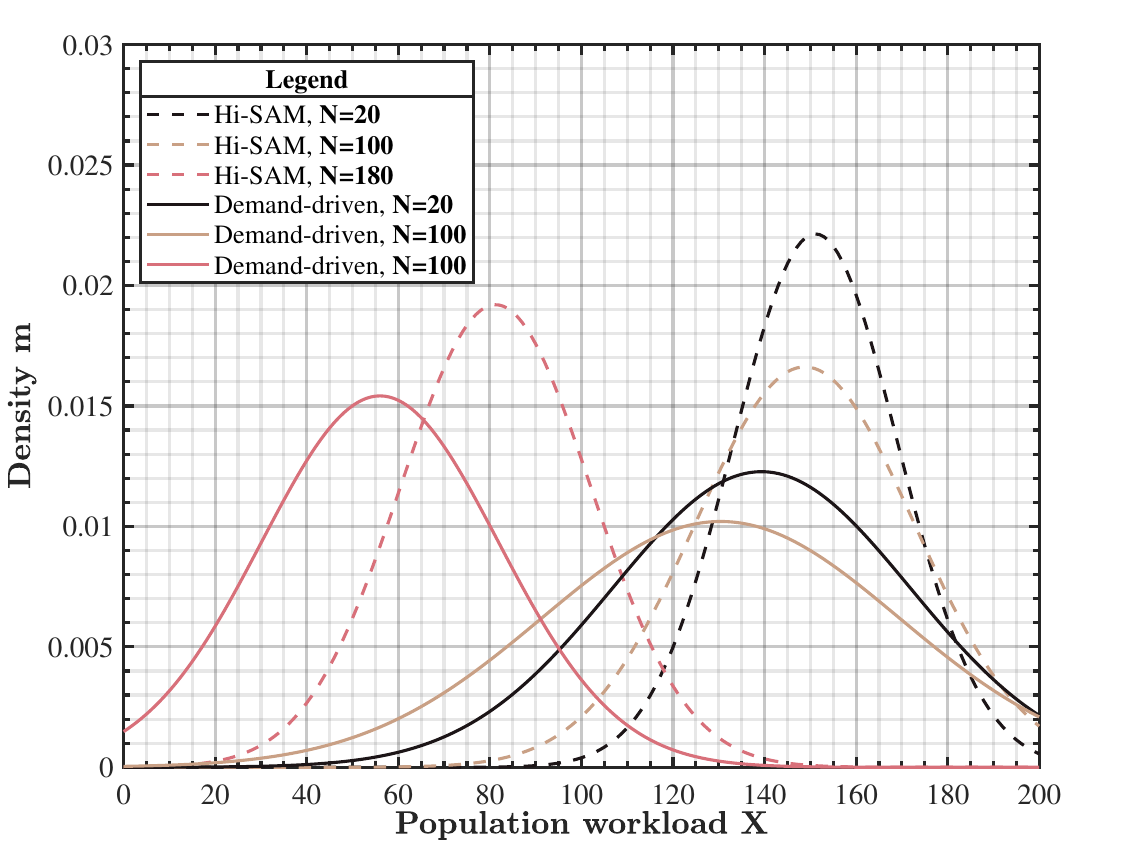}
  \vspace{-4pt}
  \caption{Workload under different population sizes.}
  \label{exp_fig_3}
  \vspace{-8pt}
\end{figure}

For wireless nodes, energy consumption is related to their lifetime. Therefore, the evaluation of their workload is critical. Fig. \ref{exp_fig_3} shows the distribution of the workload under different population sizes. Obviously, Hi-SAM has a greater mean value. This results consistent with the previous assumption made in the anomaly detection. Because an over-authentication consumes more resources, further experiments are necessary. 

\begin{figure*}[h]
  \vspace{-8pt}
\centering
\subfigure[]{\includegraphics[width=1.75in]{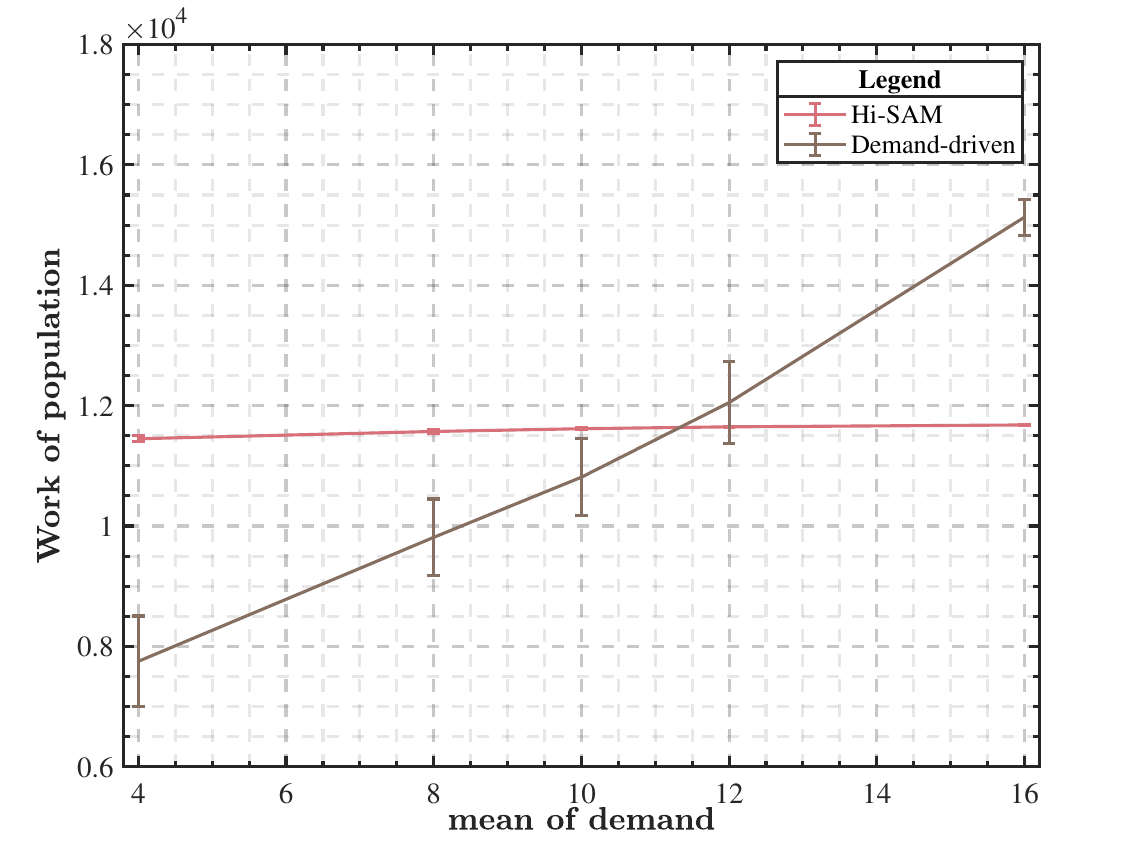}%
\hfil
\label{work-mu}}
\subfigure[]{\includegraphics[width=1.75in]{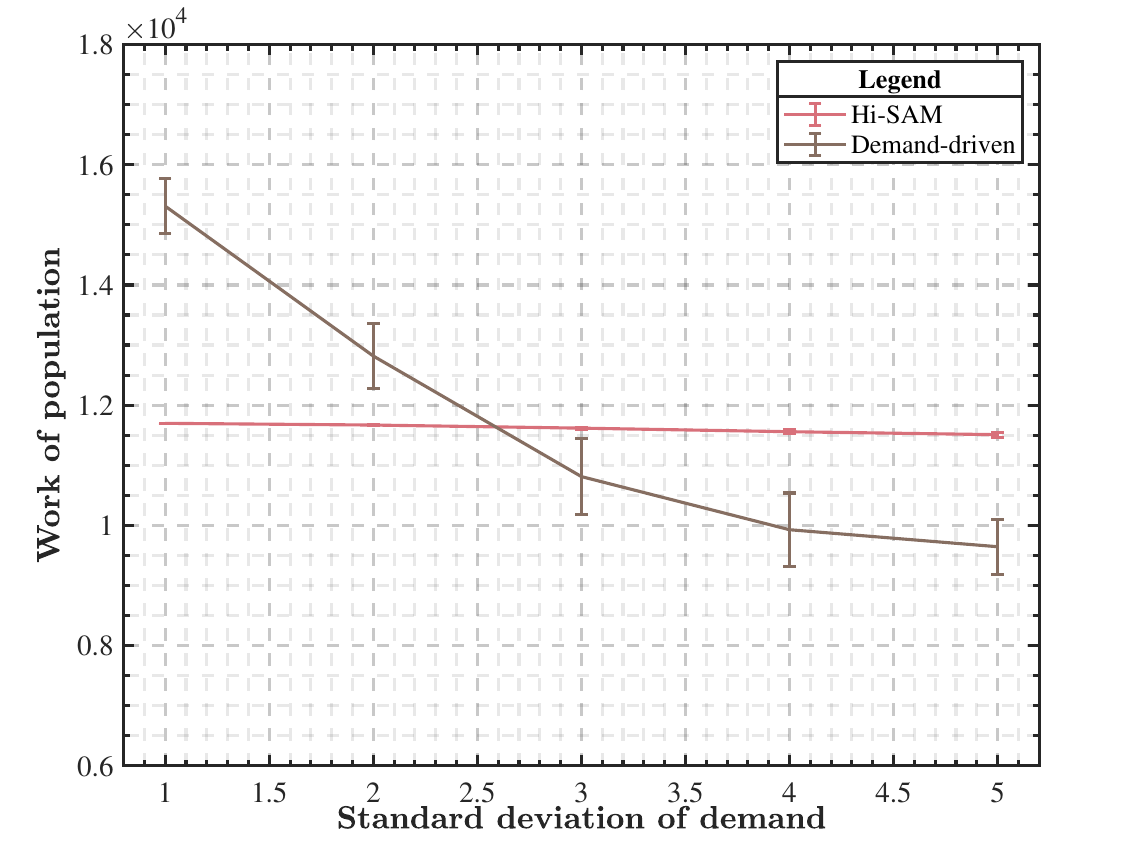}%
\label{work-sgm}}
\hfil
\subfigure[]{\includegraphics[width=1.75in]{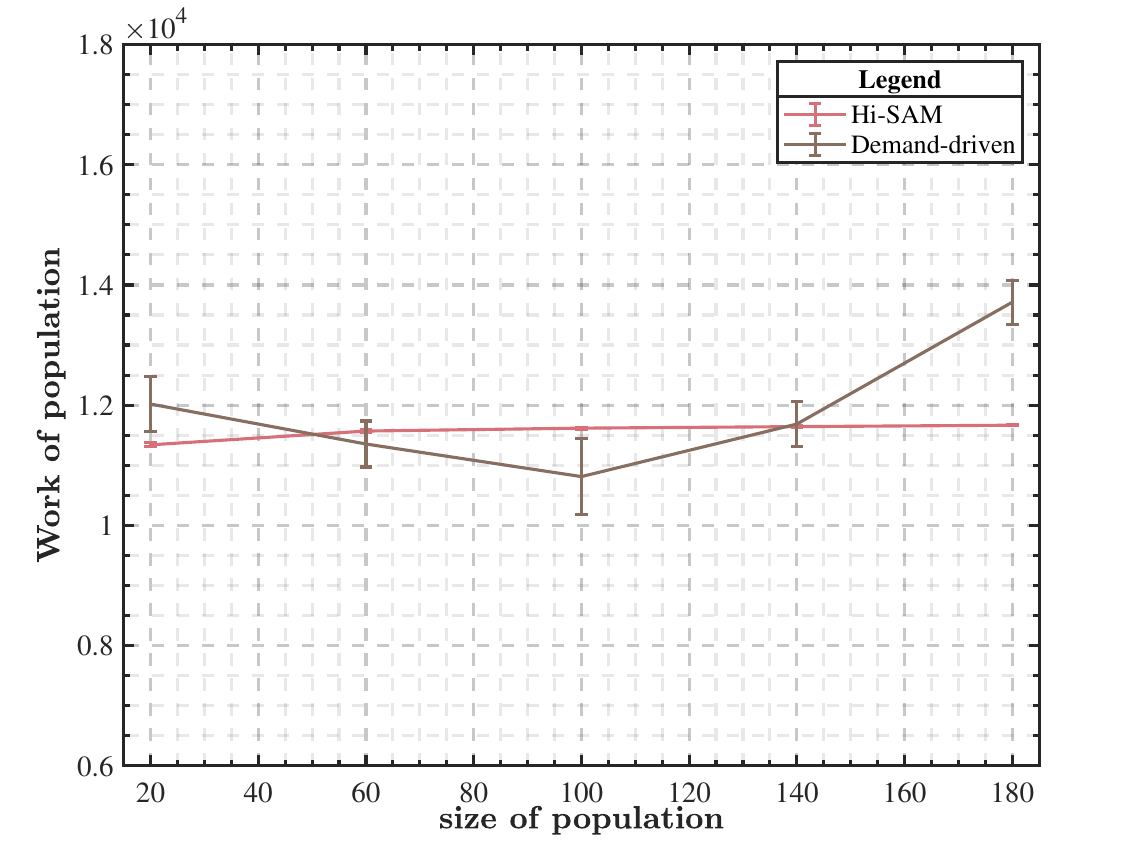}%
\label{work-N}}
\vspace{-2pt}
\caption{Comparison of workload. \\(a) Change with mean. (b) Change with variance. (c) Change with size.}
\label{exp_fig_4}
\vspace{-8pt}
\end{figure*}

From Fig. \ref{exp_fig_4}\subref{work-mu}-\subref{work-N}, the workload comparison between Hi-SAM and the demand-driven model is further studied. Firstly, a clear turning point, where $F_I=F_P/N$, is shown in Fig. \ref{exp_fig_4}\subref{work-N}. When the population size deviates from the turning point, the workload of the demand-driven model increases and exceeds Hi-SAM. Furthermore, the change of Hi-SAM is smaller and smoother. More importantly for $AP$s, Hi-SAM can release the authentication pressure when the population is large, and its workload is at least $10\%$ lower than the demand-driven model.

From Fig. \ref{exp_fig_4}\subref{work-mu} and \subref{work-sgm}, Hi-SAM's workload is higher in the case of smaller means and larger variances. This is all due to the increased share of low-demand individuals. In the worst case, Hi-SAM's workload is about $50\%$ higher than the demand-driven mode, however, its corresponding detection time is reduced by about $66\%$. Moreover, Hi-SAM is more significantly unaffected by distribution changes. In other words, Hi-SAM is more compatible with different scenarios.

In general, Hi-SAM is dramatically more insensitive to changes in demand distributions and population sizes. The advantage of Hi-SAM is more obvious when the population size increases, and it enhances the scalability of $AP$s.

\begin{figure*}[h]
  \vspace{-8pt}
\centering
\subfigure[]{\includegraphics[width=2.5in]{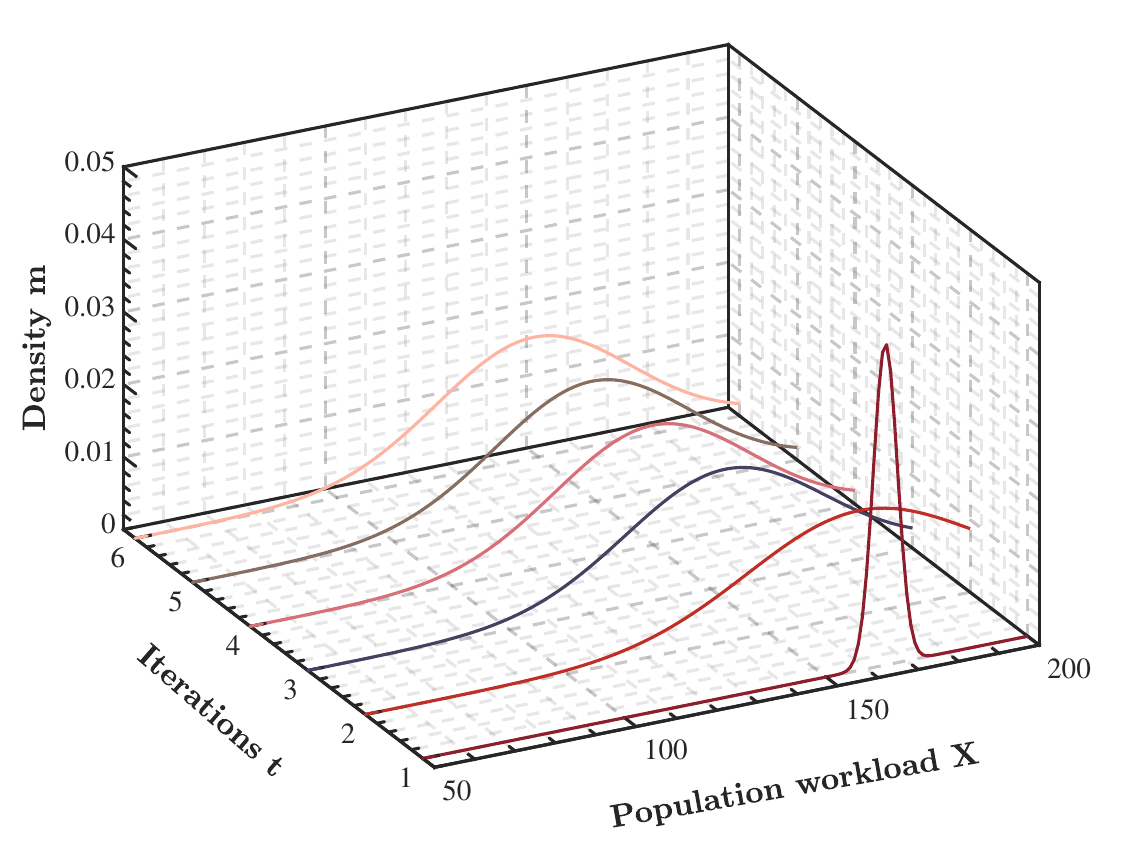}%
\label{pos_change}}
\hfil
\subfigure[]{\includegraphics[width=2.5in]{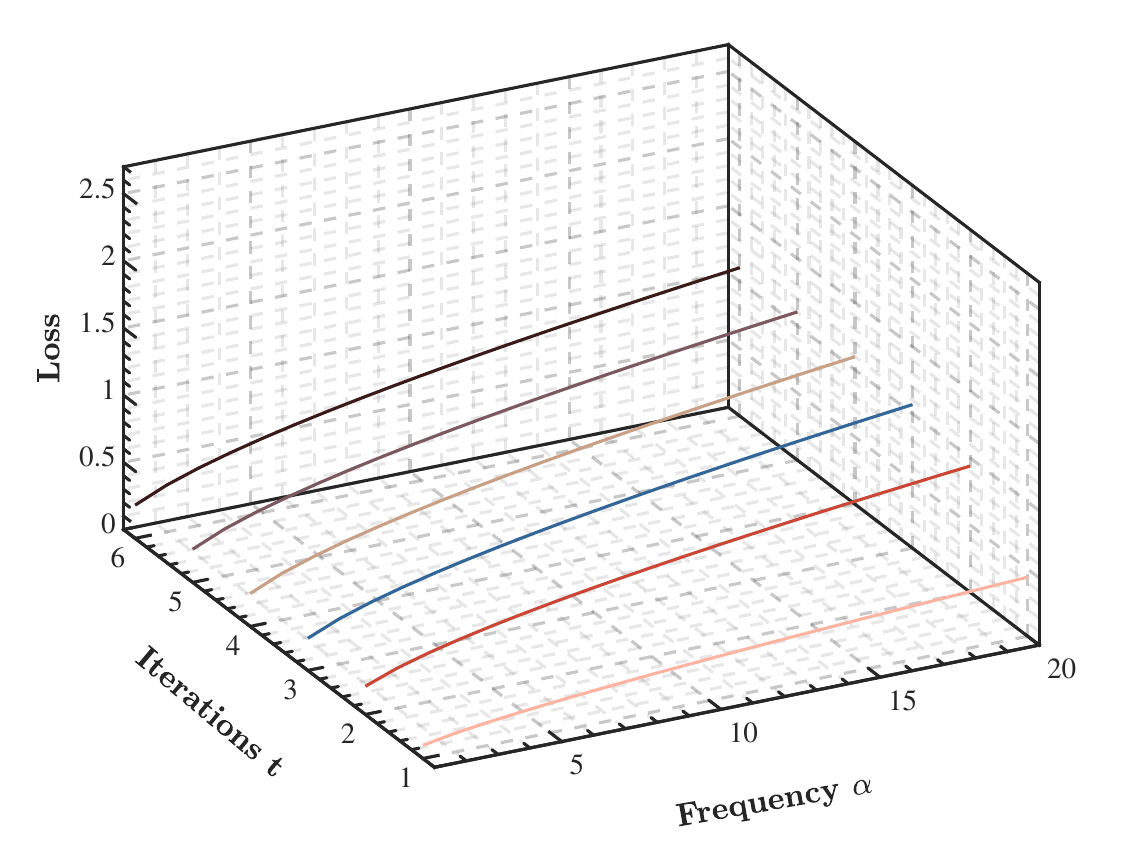}%
\label{los_change}}
\vspace{-2pt}
\caption{Process of convergence. \\(a) Change of workload distribution. (b) Change of loss.}
\label{exp_fig_6}
\vspace{-8pt}
\end{figure*}

Finally, Fig. \ref{exp_fig_6} illustrates the iterative process of Hi-SAM in the default scenario. It can be noticed that the value of the loss function varies from small to large in Fig. \ref{exp_fig_6}\subref{los_change}, the mean of the population workload varies from high to low, and the variance varies from small to large in Fig. \ref{exp_fig_6}\subref{pos_change}. In this trend, if there is an equilibrium point, the system must have a suitable step size that allows the iterations to converge, which validates the conclusions in \ref{convergence}.

\section{Conclusion}
In order to meet the demand of massive IoT devices to access the satellite network and secure the system, at the same time, to cope with the risk brought by the expansion of the attack surface, this paper designs a high-scalable authentication model for satellite-ground Zero-Trust system. Hi-SAM introduces the Proof-of-Work idea into ZTA to implement a device-differential authentication. It brings competition and dynamism to the system. Therefore, Hi-SAM use MFG to optimize authentication frequency and make competition efficient to reach consensus in a large population. MAC based DTR policy is used to to enable authentication on different frequencies and to increase the frequency of low-demand devices. The framework makes a tradeoff between the availability and security. From experiments, Hi-SAM shows greater stability and scalability in the case of large-scale devices access.

$\,$

$\,$
\bibliographystyle{IEEEtran} 
\bibliography{ref}



\end{document}